\documentclass[prb, twocolumn, groupedaddress]{revtex4}%
\usepackage{amsfonts}
\usepackage{amsmath}
\usepackage{amssymb}
\usepackage{graphicx}
\usepackage{bm}
\usepackage{subfigure}
\usepackage{color}
\usepackage[colorlinks=true,linktocpage=true,breaklinks=true]{hyperref}%
\providecommand{\U}[1]{\protect\rule{.1in}{.1in}}
\begin{document}
\title{Electrically tunable Kondo effect as a direct measurement of the chiral
anomaly in disorder Weyl semimetals }
\author{Yan-Yan Yang$^{1}$}
\author{Ming-Xun Deng$^{1}$}
\email{dengmingxun@scnu.edu.cn}
\author{Hou-Jian Duan$^{1}$}
\author{Wei Luo$^{2}$}
\author{Rui-Qiang Wang$^{1}$}
\email{wangruiqiang@m.scnu.edu.cn}
\affiliation{$^{1}$Guangdong Provincial Key Laboratory of Quantum Engineering and Quantum
Materials, GPETR Center for Quantum Precision Measurement, SPTE, South China
Normal University, Guangzhou 510006, China}
\affiliation{$^{2}$School of Science, Jiangxi University of Science and Technology, Ganzhou
341000, China}

\begin{abstract}
We propose a mechanism to directly measure the chiral anomaly in disorder Weyl
semimetals (WSMs) by the Kondo effect. We find that in a magnetic and electric
field driven WSM, the locations of the Kondo peaks can be modulated by the
chiral chemical potential, which is proportional to $\mathbf{E}\cdot
\mathbf{B}$. The Kondo peaks come from spin fluctuations within the
impurities, which apart from the temperature, relate closely to the host's
Fermi level. In WSMs, the chiral-anomaly-induced chirality population
imbalance will shift the local Fermi levels of the paired Weyl valleys toward
opposite directions in energy, and then affects the Kondo effect.
Consequently, the Kondo effect can be tunable by an external electric field
via the chiral chemical potential. This is unique to the chiral anomaly. Base
on this, we argue that the electrically tunable Kondo effect can serve as a
direct measurement of the chiral anomaly in WSMs.
The Kondo peaks are robust against
the disorder effect and therefore, the signal of the chiral anomaly survives
for a relatively weak magnetic field.

\end{abstract}
\maketitle

\section{introduction}

Weyl semimetals (WSMs), as a class of novel quantum states of matter, have
recently spurred intensive and innovative research in the field of condensed
matter
physics\cite{RevModPhys.90.015001,Zhang:2016aa,Liu864,Xiong413,Zhang:2017aa,PhysRevLett.119.136806,Zhang1594094,PhysRevLett.122.036601,PhysRevLett.123.206601,PhysRevA.92.013632,PhysRevResearch.1.033102}%
. In WSMs, the conduction and valence bands touch near the Fermi level at
certain discrete momentum points, around which the low-energy spectrum forms
nondegenerate three-dimensional Dirac cones. The band-touching points,
referred to as Weyl nodes, always come in pairs with opposite chiralities in
momentum space, which are protected by topological invariants associated with
the Chern flux and connected by the nonclosed Fermi-arc surface
states\cite{Volovik2003,Yang:2014aa,PhysRevB.91.121101,Kargarian8648,PhysRevB.83.205101}%
. The ultrahigh mobility and spectacular transport properties of the charged
Weyl fermions can find applications in high-speed electronic circuits and
computers\cite{Ali2014,Shekhar2015,PhysRevX.4.031035}.

The Weyl nodes and Fermi-arc surface states are regarded as the most
distinctive observable spectroscopic feature of WSMs. However, their
observation is sometime limited by spectroscopic resolutions, especially for
disorder WSMs whose spectrum and Weyl nodes could be obscured by the impurity
scattering\cite{PhysRevB.96.155141}. In real materials, defects or impurities
are unavoidable, and therefore, there is an urgency to find similar
smoking-gun features of WSMs in other ways, such as in transport measurements.
Of particular interest is the transport related to the chiral anomaly, which
refers to the violation of the separate number conservation laws of Weyl
fermions of different chiralities. Nonorthogonal electric and magnetic fields
can create a population imbalance between Weyl nodes of opposite chiralities,
the relaxation of which contributes an extra electric current to the system
and then results in a very unusual negative longitudinal magnetoresistance
(NLMR)
phenomenon\cite{Liu864,Xiong413,Zhang:2017aa,PhysRevX.5.031023,PhysRevLett.122.036601,PhysRevX.8.031002,Neupane:2014aa,Li:2015aa}%
. While it occurs for WSMs with the chiral anomaly, the observation of the
NLMR is only a necessary condition for identifying the WSM phase, but it is
not a sufficient condition, since other mechanisms, such as the weak
antilocalization\cite{PhysRevB.92.035203}, can also induce the NLMR
phenomenon. For a relative strong magnetic field, due to the Landau level (LL)
quantization, the chiral-anomaly-induced NLMR would exhibit quantum
oscillations. The quantum oscillations, superposed on the NLMR, can exclude
the weak antilocalization mechanism and so can be a remarkable fingerprint of
a WSM phase with the chiral
anomaly\cite{PhysRevLett.122.036601,PhysRevB.100.235105}. In disorder WSMs, as
the LLs could be broadened by the impurity scattering, the observation of the
quantum oscillations in NLMR depends strongly on the disorder
effect\cite{PhysRevB.99.165146}. What is more, the NLMR, as an indirect
measurement of the chiral anomaly, would, inevitably, be influenced by some
other complicated contributions. Therefore, it is highly desirable to find a
direct way to identify the chiral anomaly.

Recently, the Kondo effect in WSMs has attracted increasing
interest\cite{Andrew_PhysRevB.92.121109,Sun_PhysRevB.92.195124,Ma_PhysRevB.97.045148,Li_PhysRevB.98.075110,Lu_PhysRevB.99.115109}.
By using the variational method, Sun $et$ $al.$ studied the Kondo effect of the WSM bulk states and found that the spatial spin-spin correlation functions can be used to distinguish a Dirac semimetal from a WSM\cite{Sun_PhysRevB.92.195124}. Ma $et$ $al.$ investigated the Kondo screening of a magnetic impurity by the Fermi arc surface states of
WSMs\cite{Ma_PhysRevB.97.045148}. The correlation functions were shown to be highly anisotropic and possess the
same symmetry as the Fermi arcs. Li $et$ $al.$ addressed the Kondo
screening associated with the chiral anomaly\cite{Li_PhysRevB.98.075110}. It is found that the magnetic
susceptibility can be significantly enhanced by increasing the chirality
imbalance and tunable by the charge imbalance of the Weyl nodes.

In this paper, taking into account the Landau quantization,
we study the Kondo effect in electric and magnetic field driven WSMs. Usually,
the Kondo effect is insensitive to nonmagnetic external fields, and thus does
not response to external electric fields. However, it relates closely to the
Fermi level of the
host\cite{PhysRevB.75.045318,PhysRevB.81.235411,PhysRevB.82.155142,PhysRevB.84.165105,PhysRevB.87.075430,PhysRevB.88.035411,Deng_2016,Li_PhysRevB.98.075110}. In the presence of nonorthogonal electric and magnetic fields, the
chiral-anomaly-induced chirality population imbalance would lead to unequal
local Fermi levels for the paired Weyl
valleys\cite{PhysRevX.4.031035,PhysRevLett.122.036601,PhysRevB.100.235105}.
Instead of the external field independent chiral chemical potential in Ref. \cite{Li_PhysRevB.98.075110}, we consider a more realistic situation, where the chiral chemical potential is established by nonequilibrium processes,
so that the Kondo effect can be electrically tunable. For a fixed chiral chemical potential, our results recover the ones in Ref. \cite{Li_PhysRevB.98.075110}. By evolutions of the locations of the Kondo peaks with respect to the external fields, we can
identify if there exists the chiral chemical potential immediately. This
unique property suggests a scheme to directly observe the chiral anomaly.
Moreover, comparing with the quantum oscillations of the NLMR, the Kondo
effect exhibits less sensitive to the disorder effect, and therefore, by the
Kondo effect, the chiral anomaly remains observable for relatively weak
magnetic fields.

The rest of this paper is organized as follows. In Sec.\ \ref{Ham}, we
introduce the model Hamiltonian and derive Green's functions for the disorder
WSM and quantum impurities. In Sec.\ \ref{valley}, we calculate the
valley-dependent local equilibrium electron distribution function by a
recent-developed theory integrating the Landau quantization with Boltzmann
equation. The chiral anomaly modulated Kondo effect is discussed in
Sec.\ \ref{chiral} and the last section contains
some discussions about the
experimental realization and a short summary.

\section{Hamiltonian and Green's functions}

\label{Ham} A disorder WSM with two Weyl nodes in a magnetic field can be
described by the Hamiltonian%
\begin{equation}
H=\sum_{\chi=\pm}\int d^{3}\mathbf{r}\psi_{\chi}^{\dag}(\mathbf{r}%
)[\chi\upsilon_{\mathrm{F}}(\hat{\mathbf{p}}+e\mathbf{A)}\cdot\mathbf{\sigma
}+U(\mathbf{r})]\psi_{\chi}(\mathbf{r}), \label{eq_Hal}%
\end{equation}
where $\mathbf{\sigma}$ is the vector of Pauli matrices, $\psi_{\chi
}(\mathbf{r})=(c_{\chi\uparrow}(\mathbf{r}),c_{\chi\downarrow}(\mathbf{r}))$
is the two-component spinor at position $\mathbf{r}$ and $\hat{\mathbf{p}%
}=-i\hbar\boldsymbol{\nabla}$ is the momentum operator, with $\chi=\pm1$ being
chiralities of the Weyl nodes that are separated by a vector $\mathbf{b}%
=2k_{0}\hat{e}_{z}$. The disorder is modeled by $U(\mathbf{r})=\sum
_{j}V(\mathbf{r}-\mathbf{R}_{j})$, where $V(\mathbf{r}-\mathbf{R}_{j})$ is a
random potential. In realistic materials, the defects could possess internal
degrees of freedom, called quantum defects or impurities. When a fermion
encounters a quantum impurity, it has chance to be scattered off the impurity
via elastic collision or change its state by coupling with the impurity's
internal degrees of freedom.
The former leads to momentum relaxation of the fermions, which refers to the process that the momentum increment of electrons by external field is undone by the impurity scattering, making it possible for the system to reach a steady state. The momentum relaxation time can be related to the mean free path, namely, the distance that an electron travels before its initial momentum is destroyed. The latter usually results in inelastic scattering. If the internal state, such as charge and spin, of the impurity fluctuates with time, the impurity scattering can be phase-randomizing and then causes phase relaxation for the scattered fermions.\cite{Dattaele1997,Mahan2013}.
Specifically, we use the Anderson
model\cite{Andrew_PhysRevB.92.121109,Sun_PhysRevB.92.195124,Ma_PhysRevB.97.045148,Li_PhysRevB.98.075110,Lu_PhysRevB.99.115109,PhysRevB.81.235411,PhysRevB.84.165105,Deng_2016,Zheng2016}
to characterize the quantum impurities, i.e.,
\begin{equation}
H_{\mathrm{imp}}=\int d\mathbf{r}\sum_{i,\sigma}\left(  \varepsilon_{\sigma
}d_{\sigma}^{\dag}d_{\sigma}+\frac{U}{2}n_{\sigma}n_{\bar{\sigma}}\right)
\delta(\mathbf{r}-\mathbf{R}_{i})
\end{equation}
with $n_{\sigma}=d_{\sigma}^{\dag}d_{\sigma}$ the spin-dependent number
operator and $\bar{\sigma}\equiv-\sigma$, where $\varepsilon_{\sigma}$
represents the spin-dependent impurity level, $d_{\sigma}$ ($d_{\sigma}^{\dag
}$) denotes the electron annihilation (creation) operator and $U$ stands for
the Coulomb repulsion potential at the impurity site ($\mathbf{R}_{i}$). The
coupling between the impurities and WSM can be described by $H_{\mathrm{hyb}%
}=\sum_{i,\chi\sigma}\left[  t_{\sigma}d_{\sigma}^{\dag}c_{\chi\sigma
}(\mathbf{R}_{i})+t_{\sigma}^{\dag}c_{\chi\sigma}^{\dag}(\mathbf{R}%
_{i})d_{\sigma}\right]  $, where $t_{\sigma}$ denotes the hopping integral
between the itinerant electrons and the impurities.

Without loss of generality, we assume that the vector potential $\mathbf{A}$
lies in the $y$-$z$ plane with $\mathbf{A}=Bx(\cos\theta\hat{e}_{y}-\sin
\theta\hat{e}_{z})$, which defines the magnetic field $\mathbf{B}%
=\mathbf{\nabla}\times\mathbf{A}$. By rotating the spin quantization axis
along the direction of the magnetic field $\mathbf{B}=B\hat{e}_{r}$, we obtain
a single particle Hamiltonian for the clean WSMs
\begin{equation}
H_{\chi}(\mathbf{k}_{\Vert})=\chi\hbar\omega_{c}\left(
\begin{array}
[c]{cc}%
\ell_{B}k_{r} & -i\sqrt{2}a_{k_{\theta}}\\
i\sqrt{2}a_{k_{\theta}}^{\dag} & -\ell_{B}k_{r}%
\end{array}
\right)  , \label{eq_Hchi}%
\end{equation}
where $\omega_{c}=\upsilon_{\mathrm{F}}/\ell_{B}$ is the cyclotron frequency
and $\mathbf{k}_{\Vert}=k_{\theta}\hat{e}_{\theta}+k_{r}\hat{e}_{r}$, with
\begin{equation}
\left(
\begin{array}
[c]{c}%
\hat{e}_{\theta}\\
\hat{e}_{r}%
\end{array}
\right)  =\left(
\begin{array}
[c]{cc}%
\cos\theta & -\sin\theta\\
\sin\theta & \cos\theta
\end{array}
\right)  \left(
\begin{array}
[c]{c}%
\hat{e}_{y}\\
\hat{e}_{z}%
\end{array}
\right)
\end{equation}
and $\ell_{B}=\sqrt{\hbar/eB}$ the magnetic length. The ladder operators for
the Landau-gauge wavefunctions%
\begin{equation}
\phi_{n}(\xi)=\frac{1}{\sqrt{2^{n}n!\ell_{B}\sqrt{\pi}}}e^{-\xi^{2}/2}%
H_{n}(\xi)
\end{equation}
are defined as $a_{k_{\theta}}=\frac{1}{\sqrt{2}}(\xi+\partial_{\xi})$ and
$a_{k_{\theta}}^{\dag}=\frac{1}{\sqrt{2}}(\xi-\partial_{\xi})$, where
$\xi=x/\ell_{B}+\ell_{B}k_{\theta}$ and $H_{n}(\xi)$ are the Hermitian
polynomials. Including separation of the Weyl nodes, we can expand the spinor
in Eq. (\ref{eq_Hal}) as
\begin{equation}
\psi_{\chi\sigma}(\mathbf{r})=\sum_{\alpha\chi}\frac{e^{i(k_{\theta}-\chi
b_{\theta})\tilde{y}+i(k_{r}+\chi b_{r})\tilde{z}}}{\sqrt{L_{\theta}L_{r}}%
}\varphi_{\alpha\chi\sigma}c_{\alpha\chi\sigma}, \label{eq_phsir}%
\end{equation}
where $c_{\alpha\chi\uparrow}$ and $c_{\alpha\chi\downarrow}$ are,
respectively, the annihilation operators for spin states $\varphi_{\alpha
\chi\uparrow}=\left(  \phi_{n-1},0\right)  ^{T}$ and $\varphi_{\alpha
\chi\downarrow}=\left(  0,\phi_{n}\right)  ^{T}$, with $b_{\theta}=k_{0}%
\sin\theta$, $b_{r}=k_{0}\cos\theta$ and $\alpha=\{n,\mathbf{k}_{\Vert}\}$ as
a composite index. Substituting Eq. (\ref{eq_phsir}) into Eq. (\ref{eq_Hal})
yields%
\begin{equation}
H=\sum_{\alpha\chi}\psi_{\alpha\chi}^{\dag}\left(  h_{\alpha}^{\chi}%
\psi_{\alpha\chi}+\sum_{\alpha^{\prime}\chi^{\prime}}U_{\alpha,\alpha^{\prime
}}^{\chi,\chi^{\prime}}\psi_{\alpha^{\prime}\chi^{\prime}}\right)  ,
\end{equation}
where $h_{\alpha}^{\chi}=\langle\varphi_{\alpha\chi}|H_{\chi}(\mathbf{k}%
_{\Vert})|\varphi_{\alpha\chi}\rangle$ and the matrix elements of the impurity
potential in the momentum subspace are given by
\begin{align}
U_{\alpha,\alpha^{\prime}}^{\chi,\chi^{\prime}}  &  =\int d^{3}\mathbf{r}%
e^{i(\chi-\chi^{\prime})(b_{\theta}\tilde{y}-b_{r}\tilde{z})-i(k_{\theta
}-k_{\theta}^{\prime})\tilde{y}-i(k_{r}-k_{r}^{\prime})\tilde{z}}\nonumber\\
&  \times\frac{1}{L_{\theta}L_{r}}\langle\varphi_{\alpha\chi}|U(\mathbf{r}%
)|\varphi_{\alpha^{\prime}\chi^{\prime}}\rangle.
\end{align}
Within this representation, the coupling Hamiltonian between the impurity and
WSM becomes
\begin{equation}
H_{\mathrm{hyb}}=\sum_{\alpha\chi\sigma}\left(  t_{\alpha\chi\sigma}d_{\sigma
}^{\dag}c_{\alpha\chi\sigma}+t_{\alpha\chi\sigma}^{\dag}c_{\alpha\chi\sigma
}^{\dag}d_{\sigma}\right)
\end{equation}
with $t_{\alpha\chi\sigma}=t_{\sigma}\varphi_{\alpha\chi\sigma}$.

For simplicity, it is provided that the
elastic and inelastic scattering processes are mutually
independent. Subsequently, by using the Dyson equation, we obtain the
disorder-averaged retarded Green's function\cite{Mahan2013,supplement}
\begin{equation}
G_{\alpha,\alpha^{\prime}}^{\chi,\chi^{\prime}}(\epsilon)=\frac{1}{[G_{\alpha
}^{\chi}(\epsilon)]^{-1}-\Sigma_{\alpha}^{\chi}(\epsilon)}\delta_{\alpha
\alpha^{\prime}}\delta_{\chi\chi^{\prime}},
\end{equation}
where $G_{\alpha}^{\chi}(\epsilon)=1/(\langle\varphi_{\alpha\chi}%
|\epsilon+i0^{+}|\varphi_{\alpha\chi}\rangle-h_{\alpha}^{\chi})$ is the
impurity-free Green's function for the WSM and the effect of the impurity
scattering enters the Green's function through the self-energy $\Sigma
_{\alpha}^{\chi}(\epsilon)$. In the first Born approximation, the self-energy
can be given by\cite{supplement}
\begin{equation}
\Sigma_{\alpha}^{\chi}(\epsilon)=\langle\sum_{\alpha^{\prime}\chi^{\prime}%
}U_{\alpha,\alpha^{\prime}}^{\chi,\chi^{\prime}}G_{\alpha^{\prime}}%
^{\chi^{\prime}}(\epsilon)U_{\alpha^{\prime},\alpha}^{\chi^{\prime},\chi
}+t_{\alpha\chi}^{\dag}G_{d}^{r}(\epsilon)t_{\alpha\chi}\rangle_{c},
\label{eq_selfn}%
\end{equation}
in which $\langle\cdots\rangle_{c}$ stands for the configurational average and
$G_{d}^{r}(\epsilon)$ is the impurity retarded Green's function. The first
term in the right hand side of Eq. (\ref{eq_selfn}) originates from the
elastic electron scattering and the second term is due to exchanging particles
between the impurities and WSM. From Eq. (\ref{eq_selfn}), we can also
distinguish the intra- and intervalley relaxation times before taking
summation over the $\chi^{\prime}$ index, with $\tau_{\mathrm{intra}}$ and
$\tau_{\mathrm{inter}}$ corresponding, respectively, to $\chi^{\prime}=\chi$
and $\chi^{\prime}=-\chi$. The total momentum scattering rate is defined as
$\tau_{m}^{-1}=\frac{1}{\tau_{\mathrm{intra}}}+\frac{1}{\tau_{\mathrm{inter}}%
}$, which relaxes the system to a steady state. Since the impurity levels can
be occupied by electrons, we assume a screened Coulomb potential for the
impurities. Then, it is easy to estimate\cite{supplement} $\tau
_{\mathrm{inter}}/\tau_{\mathrm{intra}}\sim\left(  2k_{0}/k_{\mathrm{sc}%
}\right)  ^{4}\gg1$ where $2k_{0}$ is momentum distance between the Weyl nodes
and $k_{\mathrm{sc}}$ is the screening wave vector, which ensures the
emergence of an observable chiral chemical potential between the Weyl
valleys\cite{PhysRevLett.122.036601,PhysRevX.4.031035}.

The matrix elements of the impurity retarded Green's function are defined
as\cite{PhysRevB.96.155141}
\begin{equation}
G_{d\sigma\sigma^{\prime}}^{r}(t,t^{\prime})=-\frac{i}{\hbar}\theta
(t-t^{\prime})\langle\{d_{\sigma}(t),d_{\sigma^{\prime}}^{\dag}(t^{\prime
})\}\rangle\label{eq_Gdss}%
\end{equation}
with $\theta(x)$ the heaviside function. By using the Heisenberg equation of
motion, we derive the impurity retarded Green's function at an arbitrarily
impurity site as\cite{supplement}%
\begin{equation}
G_{d\sigma\sigma^{\prime}}^{r}(\epsilon)=\left(  \frac{1-\langle
n_{\bar{\sigma}}\rangle}{\epsilon-\varepsilon_{\sigma}-\Sigma_{\sigma}}%
+\frac{\langle n_{\bar{\sigma}}\rangle}{\epsilon-\varepsilon_{\sigma}%
-U-\Sigma_{\sigma}^{\prime}}\right)  \delta_{\sigma\sigma^{\prime}},
\label{eq_Gdd}%
\end{equation}
where $\Sigma_{\sigma}=\Sigma_{\sigma}^{r}-U\frac{\Sigma_{\sigma}^{<}%
}{\epsilon-\varepsilon_{\sigma}-U-\Sigma_{\sigma}^{r}-\tilde{\Sigma}_{\sigma
}^{r}}$, $\Sigma_{\sigma}^{\prime}=\Sigma_{\sigma}^{r}-U\frac{\Sigma_{\sigma
}^{<}-\tilde{\Sigma}_{\sigma}^{r}}{\epsilon-\varepsilon_{\sigma}%
-\Sigma_{\sigma}^{r}-\tilde{\Sigma}_{\sigma}^{r}}$ and the average occupation
$\langle n_{\bar{\sigma}}\rangle$ can be determined self-consistently by the
fluctuation dissipation
theory\cite{Deng_2016,PhysRevB.99.085106,Li_PhysRevB.98.075110}. The
self-energies above are given by\cite{supplement}%
\begin{align}
\Sigma_{\sigma}^{r}  &  =\frac{1}{2}\sum_{\alpha\chi}|t_{\alpha\chi\sigma
}|^{2}\frac{1}{\epsilon^{+}-\varepsilon_{\alpha}^{\chi}}\nonumber\\
\tilde{\Sigma}_{\sigma}^{r}  &  =\frac{1}{2}\sum_{\alpha\chi}|t_{\alpha
\chi\bar{\sigma}}|^{2}\left(  \frac{1}{\epsilon_{\sigma}^{+}-\varepsilon
_{\alpha}^{\chi}}+\frac{1}{\epsilon_{U}^{+}+\varepsilon_{\alpha}^{\chi}%
}\right) \nonumber\\
\Sigma_{\sigma}^{<}  &  =\frac{1}{2}\sum_{\alpha\chi}|t_{\alpha\chi\bar
{\sigma}}|^{2}\left(  \frac{1}{\epsilon_{\sigma}^{+}-\varepsilon_{\alpha
}^{\chi}}+\frac{1}{\epsilon_{U}^{+}+\varepsilon_{\alpha}^{\chi}}\right)
f_{\chi}(\varepsilon_{\alpha}^{\chi}) \label{eq_self}%
\end{align}
with $\epsilon^{+}=\epsilon+i\frac{\hbar}{2\tau_{m}}$, where%
\begin{equation}
\varepsilon_{\alpha}^{\chi}=\mathrm{sgn}(n)\hbar\omega_{c}\sqrt{2|n|+\ell
_{B}^{2}k_{r}^{2}}-\chi\hbar\upsilon_{\mathrm{F}}k_{r}\delta_{n,0}
\label{eq_LLs}%
\end{equation}
are LLs for the WSM, $\epsilon_{\sigma}=\epsilon-\varepsilon_{\sigma
}+\varepsilon_{\bar{\sigma}}$ and $\epsilon_{U}=\epsilon-\varepsilon_{\sigma
}-\varepsilon_{\bar{\sigma}}-U$. In each Weyl valley, the $n=0$ LL is chiral,
manifesting the chirality of the Weyl node, and all $n\neq0$ LLs are achiral.
The valley-dependent local equilibrium electron distribution function
$f_{\chi}(\varepsilon_{\alpha}^{\chi})$ will be derived in the next section.
\begin{figure}[ptb]
\centering
\includegraphics[width=\linewidth]{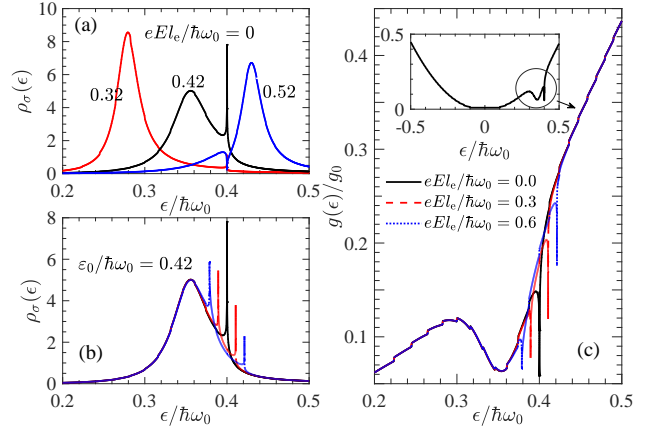}\caption{ (a)-(b) The local
DOSs $\rho_{\uparrow}(\epsilon)=\rho_{\downarrow}(\epsilon)$ at the impurity
site and (c) the DOSs $g(\epsilon)$ for the disorder-averaged WSM as functions
of energy $\epsilon$ of the itinerant electrons. The inset of (c) shows
$g(\epsilon)$ vs $\epsilon$ for $eEl_{\mathrm{e}}=0$ and $\varepsilon
_{0}/\hbar\omega_{0}=0.42$, where the region marked by the circle is replotted
in (c) for different external electric fields. For convenience, we scale the
energy and DOSs, respectively, with $\hbar\omega_{0}$ ($\sim0.01$ \textrm{eV})
and $g_{0}=(\hbar\omega_{0})^{2}/4\pi^{2}(\hbar\upsilon_{\mathrm{F}})^{3}$,
where $\omega_{0}=\upsilon_{\mathrm{F}}/\ell_{0}$ and $\ell_{0}=\sqrt
{\hbar/e[B=1\text{ }\mathrm{Tesla}]}$. Here, we set (a) $eEl_{\mathrm{e}}=0$,
$\varepsilon_{0}/\hbar\omega_{0}=(0.32,0.42,0.52)$ and (b)-(c) $\varepsilon
_{0}/\hbar\omega_{0}=0.42$, $eEl_{\mathrm{e}}/\hbar\omega_{0}=(0,0.3,0.6)$.
Other parameters are chosen as $\theta=0$, $\hbar\omega_{c}/\hbar\omega
_{0}=0.1$, $E_{\mathrm{F}}/\hbar\omega_{0}=0.4$, $\gamma\equiv\hbar/\tau
_{m}=0.05\hbar\omega_{0},t_{\sigma}=\hbar\omega_{0}$, $k_{B}T/\hbar\omega
_{0}=10^{-7}$, $\Lambda/\hbar\omega_{0}=100$ and $U/\hbar\omega_{0}=1000.$}%
\label{fig_DOSs}%
\end{figure}

\section{Valley-dependent local equilibrium electron distribution function}

\label{valley} When an external electric field $\mathbf{E}=E\hat{e}_{z}$ is
applied, the electron distribution function will deviate from the equilibrium
electron distribution function $f(\epsilon)=1/[1+e^{\beta(\epsilon
-E_{\mathrm{F}})}]$, where $\beta=1/k_{B}T$. In the relaxation time
approximation, the steady-state Boltzmann equation for the $\chi$ valley
is\cite{PhysRevLett.122.036601,PhysRevB.99.165146}%
\begin{equation}
e\mathbf{E}\cdot\boldsymbol{\upsilon}_{\alpha}^{\chi}\left(  -\frac{\partial
f_{\alpha}^{\chi}}{\partial\varepsilon_{\alpha}^{\chi}}\right)  =-\frac
{f_{\alpha}^{\chi}-f_{\chi}}{\tau_{\mathrm{intra}}}-\frac{f_{\alpha}^{\chi
}-f_{g}}{\tau_{\mathrm{inter}}}, \label{eq_Boltz}%
\end{equation}
where $\boldsymbol{\upsilon}_{\alpha}^{\chi}=\hbar^{-1}\nabla_{\mathbf{k}%
}\varepsilon_{\alpha}^{\chi}$ is the group velocity, $f_{\chi}$ and $f_{g}$
represent, respectively, the local and global equilibrium electron
distribution functions. The local equilibrium electron distribution function
equals to statistically averaging $f_{\alpha}^{\chi}$ over quantum states
around the local Fermi surface of valley $\chi$, i.e., $f_{\chi}=\langle
f_{\alpha}^{\chi}\rangle_{\chi}$ with%
\begin{equation}
\langle\cdots\rangle_{\chi}\equiv\frac{\int d\epsilon\lbrack-\partial
_{\epsilon}f(\epsilon)]\sum_{\alpha}g_{\alpha}^{\chi}(\epsilon)(\cdots)}{\int
d\epsilon\lbrack-\partial_{\epsilon}f(\epsilon)]\sum_{\alpha}g_{\alpha}^{\chi
}(\epsilon)} \label{eq_average}%
\end{equation}
and $g_{\alpha}^{\chi}(\epsilon)=-\frac{1}{\pi}\operatorname{Im}(\frac
{1}{\epsilon^{+}-\varepsilon_{\alpha}^{\chi}})$ is the momentum-resolved
density of states (DOSs) for the WSM without impurity-WSM coupling. The global
equilibrium electron distribution function $f_{g}$ can be calculated
similarly, by summation over $\chi$ separately for the numerator and
denominator in Eq. (\ref{eq_average}). Performing the local Fermi surface
average on the both sides of Eq.\ (\ref{eq_Boltz}) yields%
\begin{equation}
eE\langle\upsilon_{\alpha,z}^{\chi}\rangle_{\chi}\left(  -\frac{\partial
f_{\chi}}{\partial\varepsilon_{\alpha}^{\chi}}\right)  =-\frac{f_{\chi}-f_{g}%
}{\tau_{\mathrm{inter}}}.
\end{equation}
Together with $f_{g}=(f_{\chi}+f_{-\chi})/2$, the local equilibrium electron
distribution function can be finally obtained as%
\begin{equation}
f_{\chi}=f(\varepsilon_{\alpha}^{\chi})-eE\langle\upsilon_{\alpha,z}^{\chi
}\rangle_{\chi}\tau_{\mathrm{inter}}\left[  -\frac{\partial f(\varepsilon
_{\alpha}^{\chi})}{\partial\varepsilon_{\alpha}^{\chi}}\right]  ,
\label{eq_Local}%
\end{equation}
in which we approximated $\partial_{\varepsilon_{\alpha}^{\chi}}f_{\chi}%
\simeq\partial_{\varepsilon_{\alpha}^{\chi}}f(\varepsilon_{\alpha}^{\chi})$.
Within the framework of linear response, the valley-dependent local
equilibrium electron distribution function can be expressed as $f_{\chi
}(\varepsilon_{\alpha}^{\chi})=f(\varepsilon_{\alpha}^{\chi}+\Delta\mu_{\chi
})$, where $\Delta\mu_{\chi}=eE\langle\upsilon_{\alpha,z}^{\chi}\rangle_{\chi
}\tau_{\mathrm{inter}}$. In the absence of the magnetic field, $\langle
\upsilon_{\alpha,z}^{\chi}\rangle_{\chi}=0$ and $\Delta\mu_{\chi}$ vanishes,
while, if $B\neq0$, we can obtain a nonzero $\Delta\mu_{\chi}=-\chi\Delta\mu$,
where
\begin{equation}
\Delta\mu=eEl_{\mathrm{e}}\cos\theta\frac{1}{\int d\epsilon\left[
-\partial_{\epsilon}f(\epsilon)\right]  \varTheta(\epsilon)} \label{eq_chic}%
\end{equation}
is the so-call chiral chemical potential due to the chiral anomaly and
$l_{\mathrm{e}}=\upsilon_{\mathrm{F}}\tau_{\mathrm{inter}}$ is the intervalley
relaxation length. Here, we note
\begin{align}
\varTheta(\epsilon)  &  \equiv\sum_{\alpha}g_{\alpha}^{\chi}(\epsilon
)\nonumber\\
&  =2\sum_{n=1}^{n_{c}}\operatorname{Im}[\frac{\epsilon}{\lambda_{n}%
(\epsilon^{+})}\frac{1}{\pi}\ln\frac{\lambda_{n}(\epsilon^{+})-\Lambda
}{\lambda_{n}(\epsilon^{+})+\Lambda}]+1
\end{align}
for brevity, in which $\lambda_{n}(\epsilon)=\sqrt{\epsilon^{2}-2|n|(\hbar
\omega_{c})^{2}}$ and $\Lambda$ is a high-energy cutoff for the linear dispersion.

Replacing the momentum summation in Eq. (\ref{eq_self}) by an integral, the
self-energies can be further reduced to be $\Sigma_{\sigma}^{r}=-i\Gamma
_{\sigma}(\epsilon)$, $\tilde{\Sigma}_{\sigma}^{r}=-i\left[  \Gamma
_{\bar{\sigma}}(\epsilon_{\sigma})+\Gamma_{\bar{\sigma}}(\epsilon_{U})\right]
$ and
\begin{align}
\Sigma_{\sigma}^{<}  &  =-\frac{i}{2}\left[  \Gamma_{\bar{\sigma}}%
(\epsilon_{\sigma})+\Gamma_{\bar{\sigma}}(\epsilon_{U})\right] \nonumber\\
&  -\frac{1}{2\pi}\sum_{\chi}\Gamma_{\bar{\sigma}}(\epsilon_{\sigma}%
)\psi(\frac{1}{2}+\frac{\epsilon_{\sigma}-\chi\Delta\mu-E_{\mathrm{F}}}{2\pi
ik_{B}T})\nonumber\\
&  +\frac{1}{2\pi}\sum_{\chi}\Gamma_{\bar{\sigma}}(\epsilon_{U})\psi(\frac
{1}{2}+\frac{\epsilon_{U}+\chi\Delta\mu+E_{\mathrm{F}}}{2\pi ik_{B}T}),
\end{align}
where $\psi(x)$ is the digamma function and $\Gamma_{\sigma}(\epsilon
)=\sum_{\alpha}\pi|t_{\alpha\chi\sigma}|^{2}g_{\alpha}^{\chi}(\epsilon
)\theta(\Lambda^{2}-\epsilon^{2})$ is the linewidth function of the impurity
level due to the WSM-impurity coupling. \begin{figure}[ptb]
\centering
\includegraphics[width=\linewidth]{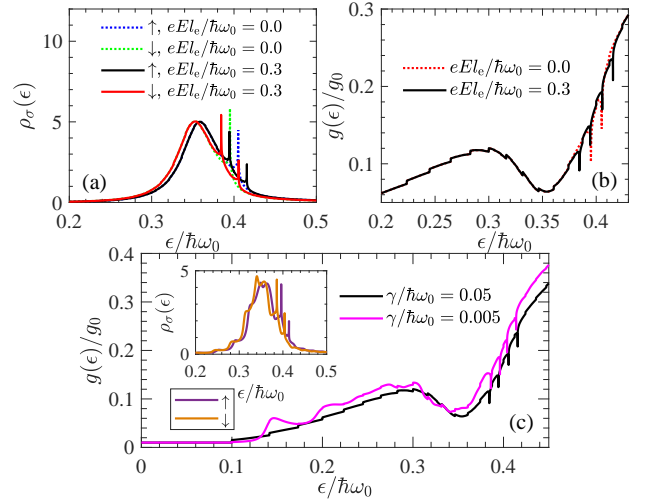}\caption{ The DOSs (a)
$\rho_{\sigma}(\epsilon)$ and (b)-(c) $g(\epsilon)$ as functions of $\epsilon$
for a finite Zeeman field $\Delta/\hbar\omega_{0}=0.0025$ at the impurity
sites, where the inset of (c) shows $\rho_{\sigma}(\epsilon)$ vs $\epsilon$
for $\gamma/\hbar\omega_{0}=0.005$. The parameters are set as (a)-(b)
$\gamma/\hbar\omega_{0}=0.05$, $eEl_{\mathrm{e}}/\hbar\omega_{0}=(0,0.3)$, and
(c) $eEl_{\mathrm{e}}/\hbar\omega_{0}=0.3$, $\gamma/\hbar\omega_{0}%
=(0.05,0.005)$. Other parameters are chosen the same as Fig. \ref{fig_DOSs}%
(c).}%
\label{fig_Gamma}%
\end{figure}

\section{Chiral anomaly modulated Kondo effect}

\label{chiral} In the following, we would consider the deep Coulomb blockade
regime, i.e., $U\rightarrow\infty$, in which we can further reduce the
impurity Green's function to a simple form
\begin{equation}
G_{d\sigma\sigma}^{r}(\epsilon)=\frac{1-\langle n_{\bar{\sigma}}\rangle
}{\epsilon-\varepsilon_{\sigma}-\Sigma_{\sigma}^{r}-\Sigma_{\sigma}^{<}}.
\end{equation}
The spin-dependent electron DOSs at the impurity site, defined as
$\rho_{\sigma}(\epsilon)=-\mbox{Im}\left[  G_{d\sigma\sigma}^{r}%
(\epsilon)\right]  /\pi$, is plotted in Figs. \ref{fig_DOSs}(a) and (b). From
Fig. \ref{fig_DOSs}(a), we can see that there appears a Lorenz peak around the
renormalized impurity level $\tilde{\varepsilon}_{\sigma}=\varepsilon_{\sigma
}+\operatorname{Re}(\Sigma_{\sigma}^{r}+\Sigma_{\sigma}^{<})$, which
characterizes the charge fluctuation between the WSM host and impurity. With
the impurity level approaching the Fermi level, an additional sharp peak
emerges to decorate the Lorenz resonance peak when the temperature is below a
critical value $T_{K}$. This sharp peak, in fact, is attributable to the Kondo
effect, which has been widely studied in varied
systems\cite{PhysRevB.75.045318,PhysRevB.81.235411,PhysRevB.82.155142,PhysRevB.84.165105,PhysRevB.87.075430,PhysRevB.88.035411,Deng_2016,Zheng2016,Andrew_PhysRevB.92.121109,Sun_PhysRevB.92.195124,Ma_PhysRevB.97.045148,Li_PhysRevB.98.075110,Lu_PhysRevB.99.115109}%
. The Kondo peak comes from the spin fluctuation at the Fermi level, which
apart from the temperature, is very sensitive to the location of the Fermi
level. In the presence of nonorthogonal electric and magnetic fields, the WSM
will exhibit the chiral anomaly, which creates a chirality population
imbalance between the Weyl valleys. The resulting chiral chemical potential
will shift the local Fermi levels of the two paired Weyl valleys in opposite
directions in energy, as shown by Eq. (\ref{eq_Local}). Consequently, in
response to the chiral chemical potential, a single Kondo peak, as seen from
Fig. \ref{fig_DOSs}(b), will split into a pair of peaks residing at the two
sides of $\epsilon=E_{\mathrm{F}}$, whose energy spacing is equal to twice of
the chiral chemical potential.
This scenario is similar to that in Ref.
\cite{Li_PhysRevB.98.075110}. The electron exchange rate $1/\tau_{\varphi}$
between the WSM and impurity increases as $\epsilon$ approaches the impurity
level, so that the DOSs of the WSM $g(\epsilon)=-\mbox{Im}\sum_{\alpha
}\mathrm{Tr}\left[  G_{\alpha,\alpha}^{\chi,\chi}(\epsilon)\right]  /\pi$, in
response to $\rho_{\sigma}(\epsilon)$, exhibits an inverse Lorenz structure,
and the Kondo peak is also observable, as indicated in the inset of Fig.
\ref{fig_DOSs}(c), where a sharp dip exists at the Fermi level. With the
electric field turned on, a single sharp dip, due to the chiral anomaly,
develops into a pair of sharp dips distributed symmetrically with respect to
the Fermi level, as shown in Fig. \ref{fig_DOSs}(c). \begin{figure}[ptb]
\centering
\includegraphics[width=\linewidth]{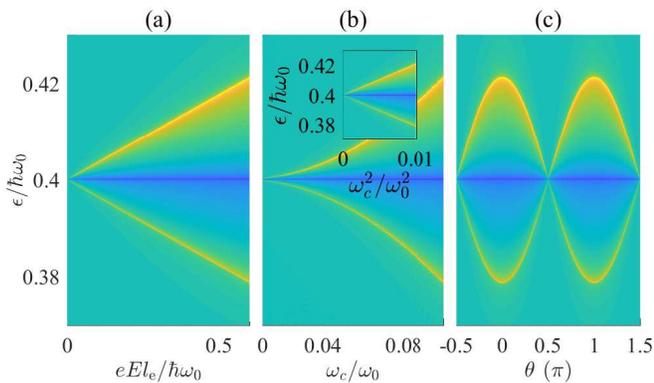}\caption{{}Evolution of (a)
$\Delta\rho_{\sigma}(E)=[\rho_{\sigma}(E)-$ $\rho_{\sigma}(E=0)]/\rho_{\sigma
}(E=0)$ with $\theta=0$ and $\hbar\omega_{c}/\hbar\omega_{0}=0.1$, (b)
$\Delta\rho_{\sigma}(B)=[\rho_{\sigma}(B)-$ $\rho_{\sigma}(B=0)]/\rho_{\sigma
}(B=0)$ with $\theta=0$ and $eEl_{\mathrm{e}}/\hbar\omega_{0}=0.6$, and (c)
$\Delta\rho_{\sigma}(\theta)=[\rho_{\sigma}(\theta)-$ $\rho_{\sigma}%
(\theta=\pi/2)]/\rho_{\sigma}(\theta=\pi/2)$ with $\hbar\omega_{c}/\hbar
\omega_{0}=0.1$ and $eEl_{\mathrm{e}}/\hbar\omega_{0}=0.6$, where the values
of $\Delta\rho_{\sigma}$ are denoted by the filled colors. The data of (b) is
replotted in the inset to show the $B$-linear ($\omega_{c}^{2}$) dependence of
the chiral chemical potential. Other parameters are the same as Fig.
\ref{fig_DOSs}(c).}%
\label{fig_EB}%
\end{figure}

The appearance of the Kondo peaks, resembling the scenario of an impurity
deposited in graphene or on the surface of topological
insulators\cite{PhysRevB.84.165105,Deng_2016}, is attributable to the
singularity of the impurity Green's function at the Fermi level. Since the
real part of the digamma function $\psi(\frac{1}{2}+\frac{\epsilon_{\sigma
}-\chi\Delta\mu-E_{\mathrm{F}}}{2\pi ik_{B}T})$ develops a sharp peak at
$\epsilon_{\sigma}-\chi\Delta\mu=E_{\mathrm{F}}$ when the temperature is lower
than the Kondo temperature $T_{K}$, there always exists a solution for
$\epsilon-\varepsilon_{\sigma}-\operatorname{Re}(\Sigma_{\sigma}^{r}%
+\Sigma_{\sigma}^{<})=0$ at $\epsilon_{\sigma}=E_{\mathrm{F}}+\chi\Delta\mu$,
which contributes a singularity to the impurity Green's function. Accordingly,
the Kondo peaks in fact develop at $\epsilon=E_{\mathrm{F}}+\chi\Delta
\mu+\sigma\Delta$, where $\Delta=\varepsilon_{\uparrow}-\varepsilon
_{\downarrow}$ is the Zeeman splitting energy of the impurity level. As it
shows, the Zeeman field on the impurity site can also result in splitting of
the Kondo peak, which is also reported in Ref. \cite{Li_PhysRevB.98.075110}.
However, in this situation, if $\Delta\mu=0$, the Zeeman field just shifts the
Kondo peaks for the two spin sectors toward different directions in energy, as
shown by the dotted lines in Fig. \ref{fig_Gamma}(a), so that each spin
component still contains only one Kondo peak. Meanwhile, the Lorenz resonance
peaks for the two spin components would separate from each other because of
broken spin degeneracy of the impurity level. Differently, the chiral anomaly
will induce a pair of Kondo peaks for both spin components, as seen from Fig.
\ref{fig_DOSs}(b) and Fig. \ref{fig_Gamma}(a), and, if $\Delta=0$, the DOSs
remain identical for the two spin species. As indicated by Fig.
\ref{fig_Gamma}(b), including both the chiral anomaly and Zeeman effect on the
impurity sites, the two Kondo dips (red-dotted line) for the WSM would split
into four dips (dark-solid line). Due to the LL quantization, the DOSs of the
WSM may exhibit quantum oscillations, which depends on the relative magnitudes
of the spacing $\Delta_{n}=\sqrt{2}\hbar\omega_{c}(\sqrt{|n+1|}-\sqrt{|n|})$
and impurity-induced broadening $\gamma\equiv\hbar/\tau_{m}$ of the LLs. The
quantum oscillations are resolvable only when $\Delta_{n}$ is much greater
than $\gamma$, and so it is expected that the quantum oscillations in the LNMR
are sensitive to the impurity scattering, especially for weak magnetic fields.
However, as shown by Fig. \ref{fig_Gamma}(c), the Kondo peaks are less
sensitive to the broadening of the LLs. As seen from the inset of Fig.
\ref{fig_Gamma}(c), the quantum oscillations also can be reflected in the
Lorenz peaks of the impurity DOSs. \begin{figure}[ptb]
\centering
\includegraphics[width=\linewidth]{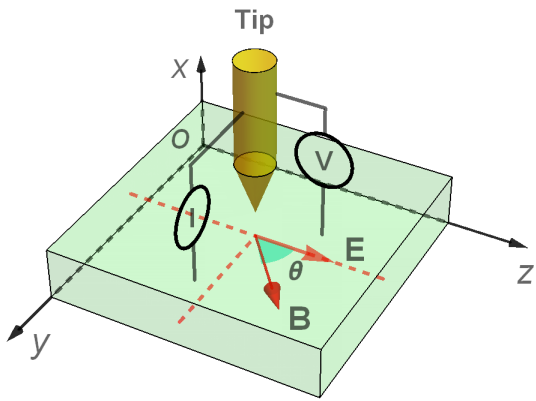} \caption{Schematics of the
device for point contact measurement\cite{Wang2016,Shuo2017,WANG2017425},
where a STM tip (brown) is attached to the top of a doped WSM slap (cyan). }%
\label{fig_setup}%
\end{figure}

To extract information of the chiral chemical potential, we plot $\Delta
\rho_{\sigma}(E)=[\rho_{\sigma}(E)-$ $\rho_{\sigma}(E=0)]/\rho_{\sigma}(E=0)$
in Fig. \ref{fig_EB}(a), $\Delta\rho_{\sigma}(B)=[\rho_{\sigma}(B)-$
$\rho_{\sigma}(B=0)]/\rho_{\sigma}(B=0)$ in Fig. \ref{fig_EB}(b) and
$\Delta\rho_{\sigma}(\theta)=[\rho_{\sigma}(\theta)-$ $\rho_{\sigma}%
(\theta=\pi/2)]/\rho_{\sigma}(\theta=\pi/2)$ in Fig. \ref{fig_EB}(c),
respectively in the $E$-$\epsilon$, $B$-$\epsilon$ and $\theta$-$\epsilon$
parameter spaces, through which the background DOSs can be subtracted to
highlight the locations of the Kondo peaks. The evolution of the energy
positions of the Kondo peaks are demonstrated by the yellow regions. The dark
blue lines along $\epsilon/\hbar\omega_{0}=0.4$ correspond to the Kondo peaks
for the case of vanishing chiral chemical potential, which locates the Fermi
energy. As seen from Fig. \ref{fig_EB}(a), for fixed $B$ and $\theta$, the
Kondo peaks will deviate from the Fermi level, with the deviation
$|\Delta\epsilon|\propto E$. For fixed $E$ and $\theta$, $|\Delta
\epsilon|\propto(\hbar\omega_{c})^{2}\sim B$, while for fixed $B$ and $E$,
$|\Delta\epsilon|\propto\cos\theta$, as indicated in Figs. \ref{fig_EB}(b) and
(c). Similar patterns also emerge in the DOSs of the WSM. This implies that
the separation of the Kondo peaks is proportional to $\mathbf{E}%
\cdot\mathbf{B}$, which demonstrates the chiral anomaly origin of the
splitting of the Kondo peaks. Therefore, the Kondo effect in magnetic and
electric field driven WSMs can capture the characteristics of the chiral
anomaly, and the observation of the electrically tunable Kondo effect can
provide an exclusive evidence for the emergence of the chiral anomaly in WSMs.

\section{Discussion and conclusion}

To date, experiments about the chiral-anomaly-modulated Kondo effect are still absent. In experiment, the chiral anomaly can be detected by using point contact spectroscopy
measurements\cite{Wang2016,Shuo2017,WANG2017425}, as depicted in Fig. \ref{fig_setup}. The setup
consists of a doped WSM slap (cyan) and a scanning tunneling microscopy (STM). The
electric and magnetic fields are applied in the $y$-$z$ plane to induce the
chiral chemical potential and the STM tip is attached to the top of the WSM
slap to measure the differential conductance between the STM and WSM.
The chemical potential of the WSM can be tuned by a gate voltage, which is not shown. For a
fixed electric and magnetic field, as the chemical potential of the
WSM varies, the differential conductance will develop a sharp peak when a local Fermi
energy encounters the renormalized impurity level. The energy locations of the
Kondo peaks correspond to the sharp peaks of the differential conductance,
whose spacing reflects the chiral chemical potential.

In conclusion, we have investigated the Kondo effect in magnetic and electric
field driven WSMs. It is found that, due to the chiral anomaly, unequal local
Fermi levels can be established between the paired Weyl valleys, and so lead
to splitting of the Kondo peaks. The external field dependent chiral chemical
potential makes the Kondo peaks electrically tunable. The electrically tunable
Kondo peaks is unique to the chiral anomaly and thus can serve as a direct
measurement of the chiral anomaly. The Kondo effect is less sensitive to the
disorder effect than transport signals, so that the chiral anomaly survives
for relatively weak magnetic fields.

\section{acknowledgements}

This work was supported by the National Natural Science Foundation of China
under Grants No. 11904107 (M.X.D), 11874016 (R.Q.W), and 11804130 (W.L), by
the Guangdong NSF of China under Grant No. 2020A1515011566 (M.X.D) and the Key
Program for Guangdong NSF of China under Grant No. 2017B030311003, GDUPS(2017)
and by the projects funded by South China Normal University under Grant No.
671215 and 8S0532.

\bibliographystyle{apsrev4-1}
\bibliography{bibKondo}

\end{document}